*How you feel about a mechanism doesn't change whether it governs you.*

# Dysmemic Pressure

*Selection Dynamics in Organizational Information Environments*

Jeremy McEntire

December 2024

## Abstract

Why do organizations comprised of intelligent individuals converge on collective delusion? This paper introduces *dysmemic pressure* as a formal mechanism explaining organizational epistemic failure. Synthesizing strategic communication theory (Crawford & Sobel, 1982), agency theory (Prendergast, 1993), and cultural evolution (Boyd & Richerson, 1985), I demonstrate how preference divergence between organizational agents generates stable equilibria where communication becomes statistically independent of reality, while transmission biases lock dysfunction into self-reinforcing states. The mechanism operates through identifiable dynamics: as the bias between sender and receiver preferences increases, communication precision degrades through progressively coarser partitions until reaching 'babbling equilibrium' where messages carry no information; simultaneously, transmission biases (content, prestige, conformity) ensure that dysfunctional signals outcompete accurate ones in the organizational meme pool. Three detailed case studies—Nokia's smartphone collapse, NASA's Challenger disaster, and Wells Fargo's account fraud scandal—illustrate the mechanism's operation across industries and failure modes. I derive five testable propositions and evaluate potential countermeasures through a mechanism design lens. The analysis reframes organizational dysfunction from moral failure to physics problem, explaining why standard interventions (culture change, leadership development, values alignment) so often fail: they treat equilibrium outcomes as behavioral problems rather than altering the selection environment that produces them.

*Keywords:* organizational behavior, information economics, cultural evolution, strategic communication, cheap talk, agency theory, organizational failure, epistemic dysfunction

## 1. Introduction

The modern organization is supposed to be an information-processing machine. Distributed knowledge flows upward through reporting structures, gets aggregated by managers, and emerges as coordinated action. The hierarchy exists, in theory, to make the whole smarter than any individual part. This premise underlies nearly a century of organizational theory, from Weber's bureaucratic rationality through Simon's bounded rationality to contemporary work on organizational learning and knowledge management.
The premise does not survive contact with observation.
Nokia's middle managers knew Symbian was failing years before the company's collapse. They sent optimistic reports anyway (Vuori & Huy, 2016). NASA's engineers knew O-rings eroded in cold weather; they signed off on the Challenger launch anyway (Vaughan, 1996). Wells Fargo's branch employees knew they were opening fraudulent accounts; they hit their quotas anyway (Independent Directors of the Board of Wells Fargo, 2017). In each case, the organization was not starved for information. It was drowning in false signals it had selected for.

The conventional explanations invoke psychology: hubris, greed, groupthink, cognitive bias. These explanations are not wrong. They are incomplete. They fail to account for the systematic nature of the failure—that is, why the same pattern recurs across industries, cultures, decades, and organizational forms. They fail to explain why organizations comprised of individually rational actors produce collectively irrational outcomes with such regularity that we can predict it.

This paper proposes a structural explanation grounded in the intersection of three literatures that have not been adequately connected: strategic communication in economics, agency theory, and cultural evolution in biology. The core argument is that organizational dysfunction of this type is the equilibrium itself, not a deviation from it. Organizations are selection environments. Ideas, reports, signals, and cultural norms compete for transmission within those environments. What survives is what is fit—regardless of whether it maps reality. And fitness, in organizational contexts, is often negatively correlated with truth.

I call this selection force *dysmemic pressure*: the systematic favoring of cultural variants that increase individual payoff while decreasing collective adaptability. The term 'dysmemic' has appeared in informal discourse on memetics (e.g., Glendinning, 2001–present) to describe the phenomenon of harmful ideas spreading rapidly. This paper provides the first formal definition of *dysmemic pressure* as an organizational selection mechanism, specifying its components and deriving testable propositions from the synthesis of strategic communication and cultural evolution theory. The terminology follows the established eu-/dys- pattern in memetics; 'eumemics' (improving meme pool quality) appears in the standard literature, while 'dysmemic' represents its logical complement, paralleling 'dysgenic' in genetics.

The contribution is not discovery, per se, but synthesis and precision—connecting literatures that have remained separate and defining the mechanism in terms that permit testable prediction. By integrating game-theoretic models of strategic communication with agency theory and cultural evolutionary models of transmission dynamics, we can describe the mechanism with sufficient clarity to identify what conditions produce it, why it is stable, why standard interventions fail, and what alternative architectures might resist it. That precision matters because it shifts the frame. Organizational dysfunction stops being a moral failing and becomes a physics problem. Physics problems do not respond to exhortation. They respond to engineering.

The paper proceeds as follows. Section 2 reviews the relevant literatures in strategic communication, agency theory, cultural evolution, and organizational failure. Section 3 develops the theoretical framework, formally defining dysmemic pressure and deriving its properties. Section 4 presents three detailed case studies illustrating the mechanism's operation. Section 5 derives testable propositions. Section 6 evaluates potential countermeasures through a mechanism design lens. Section 7 discusses implications and limitations. Section 8 concludes.

## 2. Theoretical Foundations

### 2.1 Strategic Communication and the Partition Theorem

The game-theoretic foundation comes from Crawford and Sobel's (1982) seminal paper on strategic information transmission, which formalized the conditions under which communication conveys information in the presence of conflicting interests. In their model, a Sender observes the true state of the world $t$ and transmits a costless message $m$ to a Receiver, who then takes an action y affecting both parties. The crucial variable is bias—the divergence between what the Sender wants and what the Receiver wants.

Their central result transformed how economists think about communication: as bias increases, the precision of information transmission decreases. When interests are perfectly aligned, full revelation is possible. As they diverge, communication becomes increasingly coarse, partitioning the state space into ever-larger bins. When bias becomes sufficiently large, the system collapses into what they call a 'babbling equilibrium'—a state where the Sender's messages are statistically independent of the true state, and the Receiver rationally ignores them entirely.

The babbling equilibrium is not a failure of the model. It is a prediction. Given sufficient preference divergence, silence or deception becomes the rational choice. The equilibrium is stable because no player can profitably deviate: the Sender gains nothing from truthful revelation (since the Receiver ignores messages anyway), and the Receiver gains nothing from attending to messages (since they contain no information).

Subsequent work has extended this framework considerably. Kamenica and Gentzkow (2011) developed Bayesian persuasion, showing how Senders can strategically design information structures to influence Receivers even when both parties are fully rational. Their key insight is that the Sender benefits from commitment—the ability to pre-specify a signal structure before observing the state. This has direct organizational applications: formal reporting systems, metrics, and dashboards can be understood as commitment devices that constrain the information structures available to agents.

The organizational application is immediate. Consider a manager who needs the project status report to be positive (to avoid hard conversations, protect headcount, preserve their own position) and an engineer who needs to deliver an honest assessment. Their preferences diverge. The engineer learns that accurate reports invite unwanted attention, while vague optimism satisfies the manager. Communication becomes ritual. 'We're on track' means nothing. Status meetings become noise. The babbling equilibrium has been achieved.

## 2.2 Agency Theory and the Yes Man Problem

Prendergast (1993) extended this logic specifically to internal labor markets, providing the microeconomic foundation for why executives are systematically deceived by their subordinates. His 'Theory of Yes Men' demonstrates that when organizations rely on subjective performance evaluations—where a principal assesses an agent based on judgment rather than objective metrics—the agent faces powerful incentives to conform to the principal's prior beliefs.

The mechanism is straightforward. A subordinate discovers information relevant to a strategic question. If that information contradicts what the principal believes, reporting it creates two risks: the principal may doubt the subordinate's competence (since their conclusions differ from the 'correct' view), and the principal may audit the subordinate's reasoning, consuming political capital and creating friction. The rational subordinate, anticipating these dynamics, skews reporting toward the principal's priors.

Prendergast proves that this dynamic emerges not from sycophancy but from the structure of subjective evaluation itself. The principal relies on the subordinate for information but can also conduct an independent assessment. If audit costs are high—and in complex organizations, they almost always are—the principal will rely on priors. A rational subordinate, anticipating this, will match their reports to those priors. The 'Yes Man' is not a character flaw. He is the equilibrium output of a poorly designed incentive structure.

This creates a feedback loop of confirmation bias. The principal is surrounded by apparent agreement not because they hire sycophants, but because the incentive structure converts honest agents into sycophants. The result is an organization that becomes progressively detached from reality, as the feedback mechanisms meant to correct leadership's errors are repurposed to reinforce them.

## 2.3 The Lemon Problem of Ideas

Akerlof's (1970) 'Market for Lemons' describes how information asymmetry can cause market collapse through adverse selection. In his canonical example, used car buyers cannot distinguish good cars from bad ones ('lemons'). Sellers of good cars, unable to credibly signal quality, withdraw from the market. Only lemons remain.

The same dynamic operates in the organizational idea marketplace. Proposals, assessments, and strategic recommendations are 'goods' whose quality is costly to verify. An optimistic projection and an accurate one look identical at the moment of presentation. The principal (leadership) faces the same information asymmetry as the used car buyer: they cannot easily distinguish dysmemes from eumemes.

The Akerlof logic then follows. Accurate assessments require work—data gathering, analysis, acknowledgment of uncertainty. Optimistic dysmemes require only confidence. As the market floods with cheap-to-produce dysmemes, accurate assessments become relatively more costly. Truth-tellers, unable to credibly distinguish themselves and facing higher production costs, reduce their output. The market equilibrates at a pooling equilibrium where the typical signal is uninformative—a mixture of charlatans and the genuinely deluded, with the careful analyst having exited.

## 2.4 Cultural Evolution and Transmission Bias

Game theory explains why individuals choose to transmit false signals. It does not explain why certain falsehoods dominate, why they spread, or why they prove so resistant to correction. For that, we need cultural evolution.

Boyd and Richerson's (1985, 2005) dual-inheritance theory treats cultural information—ideas, practices, norms—as replicating units subject to selection pressures analogous to (though not identical with) those operating on genes. The critical insight is that the fitness of a cultural variant is not determined by its truth value. It is determined by its transmission properties. An idea spreads because it is easy to remember, emotionally resonant, or associated with high-status individuals—not because it accurately maps reality.

Three transmission biases are particularly relevant to organizational contexts:

*Content bias* refers to the differential transmission of ideas based on their intrinsic properties. Simple, emotionally satisfying ideas outcompete complex, ambiguous ones. 'We just need to execute better' is lighter cognitive load than 'Our architecture has accumulated technical debt requiring a multi-quarter remediation effort with uncertain ROI.' The former spreads; the latter dies in the meeting where it was born. Henrich and Gil-White (2001) demonstrate that content biases operate largely below conscious awareness—people do not choose to prefer simpler explanations; they simply find them more memorable and transmissible.

*Prestige bias* refers to the preferential copying of ideas associated with high-status individuals. If the CEO believes the competitor is irrelevant, that belief cascades downward regardless of evidence. Subordinates adopt it not because they believe it, but because imitating the leader is a dominant strategy for advancement (Henrich & McElreath, 2003). The mechanism is self-reinforcing: those who adopt the prestige figure's beliefs advance, becoming prestige figures themselves and further propagating the belief.

*Conformity bias* refers to the disproportionate adoption of common beliefs. Once a belief reaches critical mass, deviation becomes costly. If everyone reports green, reporting red marks you as the problem. The pressure to conform locks in whatever belief happened to reach threshold first (Boyd & Richerson, 1985). Importantly, conformity bias operates on perceived rather than actual consensus—if people believe everyone else believes X, they will adopt X even if private dissent is widespread. This creates informational cascades where public consensus increasingly diverges from private belief (Bikhchandani, Hirshleifer, & Welch, 1992).

Blackmore (1999) and Dennett (1995) emphasize that meme fitness is orthogonal to truth—a point frequently misunderstood. The claim is not that false ideas spread better than true ones (though sometimes they do). The claim is that transmission properties and truth value are independent variables. An idea can be true and highly transmissible, true and poorly transmissible, false and highly transmissible, or false and poorly transmissible. Selection operates on transmissibility. Truth is along for the ride, or not.

## 2.5 Organizational Failure and Epistemic Dysfunction

A substantial literature documents organizational failures attributable to information pathologies, though rarely with the formal mechanism proposed here.

Vaughan's (1996) study of the Challenger disaster introduced 'normalization of deviance'—the gradual desensitization to risk as small deviations become routine. Each successful launch with O-ring erosion made the next launch with erosion more acceptable, until catastrophic failure became, in retrospect, inevitable. Vaughan explicitly frames this as a structural rather than individual phenomenon: 'No fundamental decision was made at NASA to do evil... Rather, a loss of insight was actually facilitated by... the institutional structures that monitored risk.'

Janis's (1972) groupthink model identifies conditions under which cohesive groups make defective decisions: illusions of invulnerability, collective rationalization, stereotypes of out-groups, self-censorship, illusions of unanimity, and direct pressure on dissenters. While influential, the model operates at the group level and does not explain why these dynamics persist or why they prove resistant to intervention.

Nguyen's (2020) work on epistemic bubbles and echo chambers provides a useful distinction. Epistemic bubbles arise from omission—relevant voices are simply absent. Echo chambers arise from active discrediting—outside voices are present but dismissed as untrustworthy. Organizations can exhibit both: some information never reaches decision-makers (bubble), while information that does reach them is filtered through loyalty tests that dismiss uncomfortable sources (chamber).

Levy (2022) argues that 'bad beliefs' are often rational responses to corrupted epistemic environments rather than individual cognitive failures. This perspective aligns with the mechanism proposed here: if the selection environment rewards certain beliefs regardless of truth, holding those beliefs is individually rational even when collectively catastrophic.

## 3. Theoretical Framework

### 3.1 The Synthesis

The synthesis connects these literatures through a simple observation: organizations are both strategic environments (where agents with divergent preferences communicate) and cultural environments (where ideas compete for transmission). The game theory explains stability—why no individual can profitably deviate from the dysfunctional equilibrium. The cultural evolution explains spread—why the dysfunction saturates the organization and proves resistant to correction.

Consider an organization where accurate risk reporting would benefit the collective (avoiding disasters, enabling adaptation) but harm individual reporters (inviting scrutiny, creating conflict, marking oneself as a problem). The strategic communication literature tells us this preference divergence will degrade information quality. The cultural evolution literature tells us that whatever false signals emerge will spread through content bias (simple narratives over complex ones), prestige bias (adopting the beliefs of successful people, who succeeded partly by not reporting risks), and conformity bias (reporting what everyone else reports).
The combination produces a ratchet effect. False signals crowd out true ones because they are individually advantageous. Once established, they become culturally dominant through transmission biases. Cultural dominance makes deviation even more costly (now you're fighting consensus, not just reporting bad news). The equilibrium is self-reinforcing and self-protecting.

### 3.2 Formal Definition

I define dysmemic pressure as follows:

> **Definition:** *Dysmemic pressure* is the selection force in an organization that favors cultural variants (ideas, signals, practices) which increase individual payoff while decreasing collective adaptability—that is, where internal fitness is negatively correlated with external fitness.

The relationship among these forces can be expressed as a conceptual schema:

$$\text{Dysmemic Pressure} \propto f(\text{Incentive Divergence}, \text{Transmission Ease}, \text{Verification Cost})$$

This is not a derived formula but a heuristic summarizing the mechanism's structure: pressure intensifies as any component increases. The three terms capture the distinct contributions of each theoretical foundation. *Incentive Divergence* is the Crawford-Sobel bias parameter $b$: the degree to which internal fitness (what advances individual careers) diverges from external fitness (what benefits the organization). When these are negatively correlated—when the behaviors that help the individual harm the collective—incentive divergence is high and communication degrades toward babbling equilibrium. As $b$ increases, communication precision decreases. *Transmission Ease* captures content bias from cultural evolution: simpler, more emotionally satisfying signals spread faster regardless of accuracy. *Verification Cost* captures the Akerlof asymmetry: the higher the cost to verify signal quality, the greater the adverse selection pressure on the organizational idea market.
Several components require elaboration. *Internal fitness* refers to the expected benefit to the carrier within the organization: promotion probability, conflict avoidance, status maintenance, resource allocation, job security. These are the payoffs that shape individual behavior in the Crawford-Sobel framework. *External fitness* refers to the expected benefit to the organization as a whole in its external environment: market adaptation, risk management, competitive positioning, survival. These are the outcomes that organizational design is nominally meant to optimize.
*Negative correlation* is the key condition. Internal and external fitness can be positively correlated (when rewarding truth-telling), uncorrelated (when rewards are orthogonal to information quality), or negatively correlated (when rewards punish truth-telling). Dysmemic pressure exists when the correlation is negative—when the behaviors that benefit individuals harm the organization.
'Dysmemes' are cultural variants that satisfy this condition: they help the carrier (get promoted, avoid conflict, maintain status) while harming the host organization (misleading strategy, masking risk, preventing adaptation). They outcompete 'eumemes'—truth-tracking variants—because the selection environment rewards the former and punishes the latter.
The result is not merely that individuals lie. The organizational ecology shifts. Truth-tellers exit or go silent. The meme pool becomes saturated with dysmemes. The organization loses

715f9db13715eb56a6f

the capacity to perceive reality accurately—not because individuals are stupid, but because the smart move is to participate in the collective hallucination.

### 3.3 Conditions and Dynamics

Dysmemic pressure intensifies under identifiable conditions:
*Preference divergence:* The greater the gap between what agents want and what principals want, the greater the pressure. In Crawford-Sobel terms, larger bias produces coarser communication partitions and increases the probability of babbling equilibria. Organizations where managers are evaluated on metrics that diverge from organizational health (quarterly numbers vs. long-term value, headcount vs. capability, activity vs. outcomes) exhibit greater preference divergence.
*Evaluation coupling:* When the consumer of information is also the evaluator of the producer, pressure intensifies. The engineer reporting to the manager who controls their performance review faces different incentives than the engineer reporting to an independent quality function. Decoupling evaluation from information consumption reduces bias in the Crawford-Sobel sense.
*Transmission structure:* Steep hierarchies with few horizontal connections amplify prestige bias (information flows through few high-status nodes) and reduce the error-correction capacity of distributed networks. Flat structures with many horizontal connections reduce prestige concentration but may amplify conformity bias if consensus norms are strong.
*External feedback delay:* When consequences of dysfunction are distant in time or attribution, dysmemic equilibria persist longer. Organizations in fast-feedback environments (trading desks, emergency services) exhibit less dysmemic pressure than those in slow-feedback environments (strategic planning, R&D) because reality corrections arrive before dysfunction saturates the culture.
*Exit costs:* When truth-tellers cannot easily leave, they face a choice between silence and punishment. High exit costs (specialized skills, geographic constraints, unvested compensation) increase the proportion of the population that chooses silence, accelerating dysmemic saturation. Industries with high mobility exhibit less dysmemic pressure than those with golden handcuffs.

## 4. Case Studies

Three cases illustrate dysmemic pressure operating across different industries, failure modes, and organizational contexts. Each demonstrates the mechanism's key features: preference divergence generating babbling equilibria, transmission biases locking in dysfunction, and the self-reinforcing nature of the resulting state.

### 4.1 Nokia: Fear-Induced Babbling

Nokia's collapse from mobile phone dominance to irrelevance between 2007 and 2013 represents one of the most studied failures in business history. Vuori and Huy's (2016) detailed study, based on 76 interviews with Nokia managers and engineers, documents a systematic information pathology consistent with dysmemic pressure.
The technical facts were known. Engineers understood that Symbian, Nokia's operating system, could not compete with iOS and Android. Internal assessments documented the gaps. Middle managers were aware that development timelines were unrealistic and that the organization was falling further behind with each quarter. The information existed within the organization.

It did not reach decision-makers in usable form. Vuori and Huy document a fear-based communication breakdown: 'Top managers were afraid of the external environment and made middle managers afraid of them. Middle managers were afraid of top managers and made their subordinates afraid of them.' The result was systematic upward distortion. Bad news was softened, delayed, reframed, or omitted entirely. Status reports remained optimistic long after the situation had become dire.

The mechanism fits the dysmemic framework precisely. Preference divergence was severe: top managers needed reassurance that their strategy was working; middle managers needed to avoid being identified with failure; engineers needed to avoid the scrutiny that honest assessments would invite. The Crawford-Sobel prediction follows: communication precision collapsed.

Transmission biases amplified the dysfunction. Prestige bias meant that optimistic framings endorsed by senior leaders propagated downward while pessimistic assessments from junior engineers died in the hierarchy. Conformity bias meant that once green status reports became the norm, deviation marked the deviator rather than the problem. Content bias meant that simple narratives ('we just need to execute faster') outcompeted complex ones ('our architectural assumptions are fundamentally wrong').

The equilibrium was stable. No individual could profitably deviate. An engineer who reported accurately faced career consequences without changing the outcome—the organization could not act on the information because the same dynamics suppressed it elsewhere. The rational strategy was participation in the collective delusion.

Nokia's board received consistent reassurance until the crisis was terminal. The organization did not fail for lack of information. It failed because the selection environment had eliminated the information's transmission path.

## 4.2 NASA Challenger: Normalized Deviance

The Space Shuttle Challenger disaster of January 28, 1986, killed seven astronauts when an O-ring seal failed during launch. Diane Vaughan's (1996) landmark study documented the organizational dynamics that made the disaster, in her phrase, 'an accident waiting to happen.' Engineers at Morton Thiokol, the contractor responsible for the solid rocket boosters, knew O-rings were vulnerable to cold temperatures. They had documented erosion on previous flights. The night before launch, engineers formally recommended against launching in the forecast cold conditions. They were overruled.

The dysmemic mechanism is visible in Vaughan's account of 'normalization of deviance.' Each successful flight with O-ring anomalies made the next anomaly more acceptable. The baseline shifted. What began as a concerning deviation became expected variation, then normal operation. The cultural transmission path is clear: interpretations that permitted continued launches were adopted (content bias toward launch-supporting narratives), endorsed by program leadership (prestige bias), and became consensus (conformity bias). Interpretations that would have grounded the fleet faced the opposite selection environment.

Preference divergence was structural. NASA operated under intense schedule pressure from Congress, the White House, and institutional competition. Program managers were evaluated on launch cadence. Engineers were embedded in organizations that needed launches to continue. The communication channel between technical assessment and launch decision was systematically biased toward launch.

The night before launch, when Thiokol engineers recommended delay, NASA's response is instructive. Larry Mulloy, the solid rocket booster project manager, asked Thiokol to reconsider. Thiokol's management held an off-line caucus, during which senior vice president Jerry Mason reportedly said to Robert Lund, the VP of Engineering: 'Take off your engineering hat and put on your management hat.' Lund reversed his position. The launch proceeded.

The phrase captures the dysmemic dynamic: engineering truth (O-rings fail in cold) versus management fitness (launches maintain schedule, budget, and careers). When forced to choose, the individual chose internal fitness over external accuracy. This was not a moral failure. It was a predictable response to a selection environment that had been selecting for exactly this behavior for years.

The Rogers Commission that investigated the disaster famously concluded that NASA's 'decision-making culture' had become a causal factor. The language obscures the mechanism. 'Culture' suggests something atmospheric and diffuse. The reality was selection: systematic, structural, and predictable. The organization had built an environment where launch-supporting signals were fit and launch-delaying signals were not. The culture was the output of that selection, not its cause.

### 4.3 Wells Fargo: Institutionalized Fraud

Between 2002 and 2016, Wells Fargo employees opened approximately 3.5 million accounts without customer authorization. The fraud was not hidden—it was incentivized, measured, and managed. The Independent Directors' Report (2017) documents an organizational system that selected for exactly the behavior it nominally prohibited.

The 'cross-sell' strategy required employees to sell multiple products to each customer. Performance was measured by accounts opened, with aggressive quotas tied to compensation and job security. Employees who met quotas were rewarded; those who did not were terminated. The system created a simple optimization problem: open accounts or lose your job.

The preference divergence is stark. Wells Fargo's stated objective was customer relationships generating legitimate revenue. Individual employees' objective was survival, which required meeting quotas regardless of customer consent. The gap between stated and revealed preferences was the selection environment.

The cultural transmission followed predictable patterns. New employees learned quickly what actually mattered. Training materials emphasized ethics; peer behavior demonstrated that ethics was subordinate to numbers. Managers who met quotas were promoted, becoming prestige figures whose methods were copied. Conformity pressure reinforced the behavior—teams that opened unauthorized accounts created norms that made non-participation conspicuous.

Complaints existed at every level. The company's internal ethics hotline received reports. Regional managers raised concerns. The pattern was documented in HR files and legal settlements. The information was not absent; it was systematically discounted, attributed to bad actors rather than bad systems, treated as implementation failure rather than design consequence.

The equilibrium persisted for over a decade. No individual could profitably deviate—an employee who refused to meet quotas was terminated; a manager who reduced quotas faced performance reviews based on team numbers. The organization optimized for the measured objective (accounts opened) at the cost of the stated objective (customer relationships). The resulting scandal cost Wells Fargo billions in fines and settlements, executive careers, and reputational damage that persists years later.

The case illustrates dysmemic pressure in its most explicit form. The selection environment was not subtle—it was written into job descriptions, compensation plans, and termination criteria. The organization built a machine for generating fraud and then expressed surprise when fraud emerged.

## 5. Propositions

The theoretical framework generates testable propositions about organizational information environments. These are stated as directional predictions that could, in principle, be evaluated against organizational data.

**Proposition 1 (Preference Divergence):** The greater the divergence between what advances an individual's career and what benefits the organization, the lower the information content of upward communication. As the gap between internal and external fitness incentives widens, communication precision degrades toward babbling equilibrium.

This follows directly from Crawford and Sobel (1982). Testable implications include: organizations with stronger 'up or out' cultures should exhibit less accurate upward communication; roles with high job security should produce more accurate assessments than roles with precarious employment; communication from employees with outside options should be more informative than communication from those without.

**Proposition 2 (Evaluation Coupling):** When the recipient of information is also responsible for evaluating the sender, information quality decreases. Decoupling evaluation from information consumption improves signal accuracy.

This explains why organizations with independent audit functions, ombudsmen, or protected reporting channels often detect problems earlier than those without. The prediction is that organizations that structurally separate 'who needs to know' from 'who controls your career' will exhibit less dysmemic pressure in those domains.

**Proposition 3 (Process Capture):** Any organizational process whose outputs are used to evaluate participants will, over time, optimize for evaluation success rather than process purpose. The process becomes dysmemic theater.

This is a generalization of Goodhart's Law to cultural selection. Testable implications include: OKR processes that affect compensation should exhibit less strategic information than those that do not; performance reviews that determine promotion should contain less accurate information than developmental feedback with no career consequences; planning processes should become less predictive over time as participants learn to optimize for planning metrics rather than planning accuracy.

**Proposition 4 (Intervention Decay):** Interventions that change expressed norms without changing payoff structures will exhibit initial improvement followed by regression to the pre-intervention equilibrium. The rate of regression depends on the strength of the unchanged selection pressure.

This explains the consistent failure of culture change initiatives. Meta-analyses of organizational change efforts consistently report failure rates between 60 and 80 percent (Beer & Nohria, 2000). Testable implications include: values training should produce temporary behavioral changes that decay unless reinforced by incentive changes; leadership messaging should affect behavior only when accompanied by visible changes in reward and punishment; organizational culture should resist copying—transplanting practices without transplanting selection environments should produce decay toward the host environment's equilibrium.

**Proposition 5 (External Correction):** Organizations under strong dysmemic pressure can only be corrected by external shock—information or consequences from outside the selection environment. Internal reform attempts will be absorbed into the dysmemic equilibrium.

This follows from the self-reinforcing nature of dysmemic equilibria. Testable implications include: organizations that experience market corrections, regulatory interventions, or public scandals should exhibit temporary increases in information accuracy; the magnitude and duration of improvement should correlate with the severity of the shock; internal

'transformation' initiatives without external pressure should fail at higher rates than those accompanied by external forcing functions.

## 6. Countermeasures: A Mechanism Design Perspective

If dysmemic pressure is structural rather than behavioral, effective countermeasures must alter the selection environment itself rather than exhorting different behavior within the existing environment. This section evaluates potential interventions through a mechanism design lens, asking: what structures might shift the fitness landscape such that truth-tracking variants outcompete dysmemes?

### 6.1 The Failure of Exhortation

The standard intervention portfolio—culture change initiatives, leadership development, values training, psychological safety programs—treats dysmemic outcomes as behavioral problems susceptible to education and example. The framework developed here explains why these interventions consistently fail.

Consider the typical culture change initiative. Leadership announces new values. Posters appear. Training sessions explain expected behaviors. For a period, employees perform the new norms. Then, imperceptibly, old patterns reassert. The employees who most visibly adopted the new culture often turn out to be the same ones who were best at performing the old one—they simply shifted their performance to the new script.

The initiative failed not because employees are cynical. It failed because it did not alter the selection environment. The rewards still flowed to those who satisfied superiors rather than challenged them. The penalties still fell on those who surfaced problems rather than buried them. The new values were absorbed into the dysmemic ecosystem, becoming another vocabulary for signaling compliance.

This explains why you cannot copy another organization's culture. The visible artifacts—open floor plans, all-hands meetings, mission statements—can be replicated. Without changing the underlying selection environment, the transplanted forms decay on contact with the host organization's incentive structure. Amazon's 'disagree and commit' becomes 'disagree and get fired.' Google's '20% time' becomes time spent after finishing 'real' work. Netflix's 'freedom and responsibility' becomes freedom to comply with unwritten expectations.

### 6.2 Structural Countermeasures

Effective countermeasures share a common feature: they alter the payoff matrix such that truth-telling becomes a dominant or at least viable strategy. Several structural approaches merit consideration:

*Evaluation decoupling:* Separating the recipient of information from the evaluator of its source reduces the bias in the Crawford-Sobel sense. Examples include independent audit functions that report to boards rather than management, ombudsman offices with protected status, and anonymous reporting channels with credible confidentiality. The key is structural independence—not merely policy statements that can be overridden, but governance architecture that makes the independence durable.

*Prediction markets and scoring rules:* Internal prediction markets on project outcomes, market events, or organizational metrics can elicit private beliefs with proper incentives (Hanson, 2003). Proper scoring rules reward accurate probability assessments regardless of the outcome, decoupling the payoff from what the predictor wants to be true. Implementation challenges are substantial (liquidity, manipulation, interpretation), but the mechanism directly addresses the preference divergence at the core of dysmemic pressure.

*Red teams and adversarial processes:* Institutionalized devil's advocacy, where designated teams are rewarded for finding flaws, can create protected niches for truth-telling. The key is ensuring the red team's incentives genuinely align with finding problems rather than performing opposition. Red teams that are captured by the processes they're meant to challenge become dysmemic theater themselves.

*External validation requirements:* Requiring external review of key assessments (customer advisory boards for product decisions, independent technical review for engineering claims, third-party audit for financial projections) introduces information from outside the internal selection environment. External validators face different fitness landscapes and thus different selection pressures.

## 6.3 The Maintenance Problem

Any structure that counterweights dysmemic pressure faces continuous pressure toward absorption back into the dysmemic equilibrium. The red team that becomes too influential will be defunded or captured. The independent audit function that creates too much friction will see its mandate narrowed. The prediction market that surfaces too much inconvenient truth will be discontinued or gamed.

This is not paranoia; it is the selection dynamic operating on the countermeasures themselves. Ideas and structures that threaten the dysmemic equilibrium face the same fitness disadvantages as individual truth-tellers. The countermeasures must be designed not only to work initially but to resist absorption over time.

Durable countermeasures typically require three forms of independence:

*Governance independence:* Reporting lines that do not run through the functions being assessed. The audit committee reports to the board, not the CFO. The red team reports to an executive not responsible for the project being evaluated.

*Resource independence:* Budgets and staffing that cannot be reduced as retaliation for uncomfortable findings. Multi-year commitments, protected funding sources, or external support can provide this.

*Evaluation independence:* Career consequences for the countermeasure staff that do not depend on the satisfaction of those they assess. Rotating assignments, external career paths, or tenure-like protections can provide this.

Without all three, the countermeasure will likely be absorbed. With all three, maintenance is still an ongoing effort rather than a solved problem. The physics does not disappear; it can only be counterweighted.

## 7. Discussion

## 7.1 Implications

If dysmemic pressure is structural rather than personal, several implications follow for organizational theory and practice.

First, organizational dysfunction is not evidence of bad actors. The same people, in a different selection environment, would behave differently. Blaming individuals for systemic outcomes is not only unfair—it prevents diagnosis. The person who speaks up and gets punished is not more virtuous than the person who stays silent; they merely miscalculated the payoff structure. Attributing organizational failure to individual moral failure is itself a dysmeme—it spreads because it protects the system from examination.

Second, the framework explains the stubborn failure of organizational change efforts. Meta-analyses consistently find that most change initiatives fail to achieve their stated objectives. The dysmemic lens suggests this is not implementation failure but design failure: the initiatives target behavior without targeting the selection environment that produces the behavior. They are, in effect, trying to change the output without changing the function.

Third, some organizations may be beyond internal repair. When dysmemic pressure has saturated the meme pool sufficiently, the truth-tellers have already exited. The remaining population cannot recognize dysfunction because dysfunction is all they know. The culture is not mistaken about reality; it has constructed an alternative reality that is internally consistent and externally fatal. Correction requires external shock—market failure, regulatory intervention, scandal, or replacement of the organization entirely.

Fourth, external perspective is structurally necessary rather than merely helpful. An organization trapped in dysmemic equilibrium cannot validate its own outputs. The same biases that distort the information also distort the assessment of whether information is distorted. Outside observers—consultants, boards, investors, regulators—are not luxuries but requirements for organizations that wish to maintain contact with reality. This provides a functional justification for governance structures that might otherwise appear as mere overhead.

Fifth, understanding dysmemic pressure does not exempt you from it. Awareness is necessary but not sufficient. The forces remain operative. The question is whether sufficient counterweight has been built—structures that protect variance, mechanisms that surface truth, governance that maintains independence from the drift toward comfortable consensus.

## 7.2 Limitations and Boundary Conditions

The framework has important limitations that bound its applicability.

First, the mechanism operates most powerfully at scale. Small organizations with direct observation and tight feedback loops may not develop strong dysmemic pressure because the information pathologies are quickly corrected by reality. The framework applies primarily to organizations large enough that information must flow through multiple nodes and slow enough that consequences are temporally distant from actions.

Second, the framework does not address all forms of organizational failure. Failures due to external shocks, technological disruption, resource constraints, or genuinely unforeseeable events are not explained by dysmemic pressure. The mechanism applies specifically to failures where relevant information existed within the organization but was not transmitted, processed, or acted upon.

Third, the propositions are stated directionally rather than precisely. Quantifying dysmemic pressure, predicting thresholds for babbling equilibria, or specifying the functional form of intervention decay would require empirical work beyond the scope of this paper. The framework generates predictions but does not, at this stage, generate point estimates.

Fourth, the mechanism design countermeasures are evaluated conceptually rather than empirically. While the logic suggests they should be effective, real-world implementation faces challenges not addressed here: political resistance, cost constraints, unintended consequences, and the possibility that novel interventions create novel dysmemic adaptations.

Fifth, the framework takes the existence of preference divergence as given. It does not address why organizations develop structures that create such divergence in the first place, or why some organizations maintain alignment better than others. A complete theory would need to explain the origins of dysmemic selection environments, not merely their consequences.

## 7.3 Future Research

Several directions for future research emerge from the framework.

Empirical measurement of dysmemic pressure is the most pressing need. This might involve surveys measuring perceived preference divergence, content analysis of organizational communications over time, comparison of internal assessments with external outcomes, or experimental manipulation of selection environments in organizational settings. The propositions generate testable predictions; testing them requires operationalization.

Comparative organizational analysis could identify structural features associated with resistance to dysmemic pressure. Are there industries, governance forms, or organizational designs that exhibit systematically better information environments? What do they have in common? Case selection focusing on variation rather than failure might illuminate protective factors.

Intervention studies, ideally randomized or quasi-experimental, could evaluate the countermeasures proposed here. Does evaluation decoupling actually improve information quality? Do prediction markets elicit more accurate assessments than traditional reporting? How long do interventions persist before absorption? The mechanism design literature provides tools for such evaluation, but organizational contexts present implementation challenges that merit study in their own right.

Integration with adjacent literatures could enrich the framework. The psychological safety literature (Edmondson, 1999) addresses similar phenomena at the team level; connection might reveal how micro-level dynamics aggregate to organizational-level equilibria. The institutional theory literature addresses how organizational forms spread and persist; connection might explain how dysmemic selection environments themselves propagate across organizations.

## 8. Conclusion

Organizations fail not because they lack information but because they select against it. The selection is not random. It follows predictable dynamics: strategic incentives that make truth costly, transmission biases that make comfortable falsehoods sticky, conformity pressures that lock in whatever dysfunction reaches critical mass first.

I have called this selection force dysmemic pressure. The name is new. The phenomenon is ancient. Every organization that has ever collapsed while its members privately knew the collapse was coming has experienced it. Every reform that decayed back into the dysfunction it was meant to address has fallen victim to it. Every leader who has asked 'why didn't anyone tell me?' after a preventable disaster has discovered, too late, what it produces.

The contribution here is synthesis and precision. By connecting the game-theoretic literature on strategic communication to agency theory and the cultural evolution literature on transmission dynamics, we can describe the mechanism with sufficient clarity to identify what conditions produce it, why it is stable, why standard interventions fail, and what alternative architectures might resist it.

That precision matters because it shifts the frame. Organizational dysfunction stops being a moral failure and becomes a physics problem. Physics problems do not respond to exhortation. They respond to engineering. You do not convince gravity to behave differently. You build structures that account for its operation.

The question for any organization is not whether dysmemic pressure exists—it does, always, at scale. The question is whether anything counterweights it. Whether the selection environment has been deliberately designed to protect truth. Whether structures exist that reward accuracy over performance, dissent over consensus, reality over comfort.

Where such structures exist and are defended, organizations retain the capacity to adapt. Where they do not, the drift continues—imperceptible, comfortable, and ultimately fatal.